\documentclass[10pt,journal,compsoc]{IEEEtran}

\ifCLASSOPTIONcompsoc
  \usepackage[nocompress]{cite}
\else
  \usepackage{cite}
\fi

\usepackage{cite}
\usepackage{amsmath,amssymb,amsfonts}
\usepackage{graphicx}
\usepackage{listings}
\usepackage{textcomp}
\usepackage{xcolor}
\usepackage{subfigure}
\usepackage[noend]{algpseudocode}
\usepackage{algorithmicx,algorithm}
\usepackage{threeparttable}
\usepackage{tikz}
\usetikzlibrary{fit, calc}
\usetikzlibrary{decorations}
\usepackage{verbatim}
\usepackage{pgfplotstable}

\definecolor{codegreen}{rgb}{0,0.6,0}
\definecolor{codegray}{rgb}{0.5,0.5,0.5}
\definecolor{codepurple}{rgb}{0.58,0,0.82}
\definecolor{backcolour}{rgb}{0.95,0.95,0.92}

\lstdefinestyle{mystyle}{
    backgroundcolor=\color{backcolour},   
    commentstyle=\color{codegreen},
    keywordstyle=\color{magenta},
    numberstyle=\tiny\color{codegray},
    stringstyle=\color{codepurple},
    basicstyle=\ttfamily\footnotesize,
    breakatwhitespace=false,         
    breaklines=true,                 
    captionpos=b,                    
    keepspaces=true,                 
    numbers=left,                    
    numbersep=5pt,                  
    showspaces=false,                
    showstringspaces=false,
    showtabs=false,                  
    tabsize=2
}

\begin{document}

\title{Shrinking the Kernel Attack Surface Through Static and Dynamic Syscall Limitation}

\author{Dongyang Zhan*,~\IEEEmembership{Member,~IEEE,}
        Zhaofeng Yu, 
        Xiangzhan Yu,
        Hongli Zhang
        and~Lin Ye
\IEEEcompsocitemizethanks{\IEEEcompsocthanksitem D. Zhan, Z. Yu, X. Yu, H. Zhang and L. Ye are with the School of Cyberspace Science, Harbin Institute of Technology, Harbin,
Heilongjiang, 150001.\protect\\
E-mail: \{zhandy, 20S003135, yuxiangzhan, zhanghongli, hityelin\}@hit.edu.cn
\IEEEcompsocthanksitem * Corresponding Author}
}

\IEEEtitleabstractindextext{
\begin{abstract}
Linux Seccomp is widely used by the program developers and the system maintainers to secure the operating systems, which can block unused syscalls for different applications and containers to shrink the attack surface of the operating systems. However, it is difficult to configure the whitelist of a container or application without the help of program developers. Docker containers block about only 50 syscalls by default, and lots of unblocked useless syscalls introduce a big kernel attack surface. To obtain the dependent syscalls, dynamic tracking is a straight-forward approach but it cannot get the full syscall list. Static analysis can construct an over-approximated syscall list, but the list contains many false positives. In this paper, a systematic dependent syscall analysis approach, sysverify, is proposed by combining static analysis and dynamic verification together to shrink the kernel attack surface. The semantic gap between the binary executables and syscalls is bridged by analyzing the binary and the source code, which builds the mapping between the library APIs and syscalls systematically. To further reduce the attack surface at best effort, we propose a dynamic verification approach to intercept and analyze the security of the invocations of indirect-call-related or rarely invoked syscalls with low overhead.
\end{abstract}

}

\maketitle

\IEEEdisplaynontitleabstractindextext

\IEEEpeerreviewmaketitle

\IEEEraisesectionheading{\section{Introduction}\label{sec:introduction}}

\IEEEPARstart{T}{he} Linux Seccomp (short for secure computing mode) is a security module of the Linux kernels, which allows a process to filter syscalls using a configurable policy implemented with the Berkeley Packet Filter (BPF) rules. By applying it to applications, the kernel attack surface is shrank, because the syscalls that can be invoked are reduced and the vulnerabilities in these syscalls are mitigated. An important application scenario is the Docker container, which isolates sensitive syscalls from the containers to secure the host operating system. Containers are considered to be the standard for cloud native, which is one of the most important infrastructures for service computing\cite{sultan2019container}.

However, it is not easy for system administrators to leverage the Seccomp to secure the kernel, since the Seccomp configuration cannot be generated automatically. If the developers of an application did not configure the Seccomp, it is difficult for others to get the dependent syscall list. For Docker containers, the containers are allowed to invoke more than 250 syscalls by default, which leaves a big attack surface. There are some vulnerabilities (e.g., CVE-2017-7308\cite{CVE-2017-7308}, CVE-2017-5123\cite{CVE-2017-5123}, CVE-2016-8655\cite{CVE-2016-8655}) in syscalls can be exploited to perform the privilege escalation attacks from containers. Therefore, it is necessary to obtain the Seccomp configurations of applications systematically for system security.

There are two kinds of approaches to analyze the dependent system call list of a binary executable, including dynamic tracking and static analysis. Dynamic tracking is a straight-forward way to obtain the invoked syscall list by tracking the target executable. However, the problem is the dynamic approaches cannot get the full list. Incomplete list is unacceptable for cloud computing or other practical scenarios, since the Seccomp can influence the normal execution of the target application. Static approaches are able to get the full list, but they also face the challenge of false positives. Most applications invoke syscalls through the dynamically-linked libraries (e.g., glibc). Analyzing the dependent libraries can build the mapping between the library APIs and the syscalls. However, the result of static analysis has many false positives in indirect call analysis. The existing static analysis approaches can only obtain the over-approximated destinations of indirect calls, so the extra destinations that can never be the targets of indirect calls enlarge the call graph. The enlarged call graph results in a lot of extra syscalls that will never been actually involved, and the extra syscalls are considered false positives syscalls. The false positive syscalls in the over-approximate syscall list enlarge the attack surface.

To address the problems of the current static and dynamic approaches, a systematic attack surface reducer, sysverify, based on static analysis and dynamic verification is proposed, which leverages static analysis to obtain the over-approximate dependent syscall list for Seccomp configuration and dynamically verifies the suspicious syscalls introduced by the static analysis. Through the combination of static analysis and dynamic verification, our approach can achieve precise attack surface reduction at best effort for docker containers in the cloud. 

The dependent syscall list of a binary executable is obtained systematically by leveraging the static analysis of the target binary and dependent libraries. There are two ways for binary executables to invoke syscalls, and our system is able to handle both of them. For invoking syscalls via the library APIs, a mapping between the library APIs and corresponding syscalls is built through the systematic binary analysis and source code analysis.

The direct function call graph is constructed by the binary analysis. Due to the complicated implementations in source code (such as macro code, alias function names, etc) of different libraries, analyzing the binary is more systematic. The indirect function call graph is constructed by the source code analysis. We leverage a two-layer indirect call  analysis to obtain possible callees of the indirect calls. Since the source code contains more callsite information (i.e., types of parameters), using the source code analysis to identify the possible callees is more precise. Based on the combined function call graph, the mapping between the library APIs and the syscalls can be built. For the cases of invoking syscalls through embedded assembly code or the syscall() API of glibc, a static analysis approach is proposed to extract the syscall number. Based on the dependent syscall list, the Seccomp can be used to shrink the attack surface at run time. 

However, the static analysis cannot solve the false positive problem of the indirect call analysis. Some possible callees of the indirect calls identified by the static analysis are false positives, which enlarges the function call graph and the attack surface. To achieve the precise attack surface reduction, we propose a dynamic verification approach to further determine if a indirect-call-related syscall is a false positive. When an indirect-call-related syscall is invoked, we check the user-space stack to reconstruct the invocation path. By comparing the run-time invocation path with the function call graph obtained by the static analysis, we can determine if the syscall is really triggered by a secure path instead of a unknown path. Besides, many syscalls in the call graph of the static analysis are invoked rarely, verifying these syscalls can also secure the kernel.

To the best of our knowledge, this is the first work combining static analysis and run-time verification together to shrink the attack surface of the operating system at best effort. In addition, the static analysis leverages both binary analysis and source code analysis, which is more systematic compared with the approaches only analyzing the binary\cite{demarinis2020sysfilter} or source code\cite{ghavamnia2020confine}. Sysverify constructs the direct call graph by analyzing the binary, so it does not need to handle the complex source code implementation details of different shared libraries. For indirect calls, sysverify analyzes the source code, which contains more information than the binary, so the analysis is more precise.

In summary, the contributions of our paper are as follows.
\begin{itemize}
  \item An attack surface reduction approach combining static analysis and dynamic verification is proposed to limiting the accessible syscalls of the target programs. 
  \item A systematic static analysis approach is proposed to obtain the dependent syscall list of the target binary executable, which analyzes the binary to construct the direct function call graph and analyzes the source code to build the indirect function call graph.
  \item To further reduce the attack surface, a dynamic verification approach is proposed to check if the indirect-call-related or rarely invoked syscalls are really triggered by the secure paths.
  \item A comprehensive evaluation is performed to test the effectiveness, CVE mitigation and performance overhead of our approach. The results show that our approach can effectively reduce accessible syscalls for programs with very low overhead, and about 71 CVEs can be mitigated on average for each program.
\end{itemize}

The rest of this paper is organized as follows. Section \ref{background} gives the background. The system overview is described in Section \ref{sec:overview}. Section \ref{S:static-analysis} tells the design of the systematic static analysis. The dynamic verification is presented in Section \ref{s:dynamic}. Section \ref{s:implementation} describes some important details in implementation. Section \ref{s:evaluation} evaluates the effectiveness and performance of the prototype system. The related work is summarized in Section \ref{s:related-work}. Section \ref{s:conclusion} concludes this paper.  

\section{Background} \label{background}

\subsection{Linux Seccomp}
The Linux Seccomp (short for secure computing mode)~\cite{Seccomp} is a computer security facility in the Linux kernel. It allows a process filtering incoming syscalls with a configurable policy using the Berkely Packet Filter (BPF) program, which is a powerful filter in the Linux kernel and will be executed over struct seccomp\_data reflecting the system call number, arguments, and other metadata. 

There are a large number of syscalls exposed to the user processes. In recent years, there are lots of vulnerabilities found in the kernel syscalls. Exploiting the vulnerabilities in the kernel syscalls (e.g., CVE-2017-7308, CVE-2017-5123) are usually used to compromise the operating system kernel. The CVE-2017-7308 locates in the packet\_set\_ring function, which does not properly validate certain block-size data, and allows local users to cause a denial of service. The attack vector dependents on two syscalls (i.e., socket and setsocketopt). Therefore, the exposed syscalls introduce a big attack surface. If a program does not rely on this syscall in normal execution, the syscall can be blocked by Seccomp, so that the vulnerability can be mitigated. Since there are many syscalls not used by the user processes, the accessible syscall set can be reduced to shrink the attack surface. 

The Seccomp filtering provides a way for processes to specify the syscall set that can be invoked by itself. After the configuration, if a program tries to invoke other syscalls, the operating system will reject the invocation. The Linux Seccomp is widely used by programs, such as QEMU, OpenSSH, etc. However, many developers do not configure the Seccomp profiles, which is insecure for the operating systems, especially in cloud computing. If an unprotected program that provides services to the Internet is compromised by attackers, it could exploit the vulnerabilities in all of the available syscalls. Therefore, leveraging the Seccomp to secure the operating system is important to system maintainers. This paper aims to shrink the attack surface without the cooperation of developers by performing the precise syscall limitation for programs.

\subsection{Shrinking Attack Surface of Containers}
Shrinking the kernel attack surface is very important for the container scenario, since containers share the host's operating system. In the virtual machine scenario, every virtual machine has its own operating system, even though the vulnerabilities of the VM's operating system are exploited, the attacker cannot go outside the virtual machine. However, attackers in containers are able to exploit the vulnerabilities in the accessible syscalls to perform the privilege escalation attacks, so the isolation of containers is weaker than that of virtual machines. Docker uses the Seccomp to perform the syscall access control on containers. But, Docker only blocks about 50 sensitive syscalls by default, leaving a big attack surface. 

In order to narrow the range of accessible syscalls, the Nabla Container \cite{nabla} was proposed, which is a more isolated container architecture proposed by IBM. It narrows the host's attack surface by narrowing the range of the accessible syscalls. The library OS\cite{tsai2014cooperation} is used to integrate most of the syscalls into the container applications, so that the container can only access the 7 syscalls, including: read, write, exit\_group, clock\_gettime, ppoll, pwrite64 and pread64. Therefore, the Nabla Containers still rely on some services of the host operating system, including memory management, file system and network system. Through the testing, the Nabla Containers can access less kernel code compared with the Docker containers. However, all container images and applications of the Nabla containers need to be recompiled, so it is not easy to promote the Nabla Container.

There are some approaches proposed to further limit the range of syscalls that the Docker containers can access. SPEAKER\cite{lei2017speaker} divides the accessible syscalls into two categories: the short-term access syscalls and the long-term access syscalls. It dynamically updates the Seccomp strategy. when the container is started, only short-term access syscalls are allowed; when the container is running, only long-term access syscalls are allowed. \cite{wan2017mining} is also an access control method based on the Seccomp, which dynamically tracks the invoked syscalls of the process to establish a set of syscalls that may be accessed, then it uses this set as a whitelist to directly restrict the syscall access of the container. \cite{barlev2016secure} can automatically configure the Seccomp based on a dynamic learned syscall set. However, these methods require a long time to dynamically build the syscall collections, and they cannot build a comprehensive collection, which is not acceptable for practice. 

Besides the dynamic tracking, Confine\cite{ghavamnia2020confine} leverages static analysis to build the mapping between APIs of the dependent libraries and syscalls, then it generates the syscall list of the target program according to its API invocation. Sysfilter\cite{demarinis2020sysfilter} and \cite{zeng2014tailored} analyze the binary to obtain the over-approximate syscall set. However, static analysis cannot analyze the indirect calls precisely, which introduces many false positive to the function call graph and the API-syscall mapping. Since the Seccomp configuration is based on the mapping, the static analysis approaches enlarge the attack surface. In addition, only analyzing the binary or the source code to build the mapping is not systematic due to the complexity of source code implementation of different shared libraries and the lack of callsite information of indirect calls in the binaries.

In this paper, we aim to overcome the shortcomings of the static analysis approaches by proposing a systematic analysis approach to build the API-syscall mapping and a dynamic verification approach to further shrink the attack surface at best effort.

\section{Design Overview}
\label{sec:overview}

In this section, we first introduce the observation. Based on the observation, we introduce the architecture of our approach at a glance. 

\subsection{Observation}
To achieve the precise automatic Seccomp configuration, the precise mapping between the library APIs and syscalls should be constructed, since most programs leverage the dynamically-linked libraries to invoke syscalls. Fortunately, the execution environment of the target programs is usually managed by the cloud service providers and most dynamically-linked libraries are open-source, making this task possible. A program could use many libraries, and a library may use the APIs of other libraries. Although some developers leverage the self-written dynamic link libraries in their programs, most public or self-written libraries use the glibc to invoke syscalls. Therefore, we can analyze the binaries of the dependent self-written libraries to find out which APIs of glibc they depend on. Based on the mapping between APIs and syscalls of glibc, the dependent syscalls of the self-written libraries can be analyzed. If the libraries invoke syscalls directly, the dependent syscalls can be obtained through binary analysis.

For containers, most cloud tenants execute their programs based on the public images from the Docker Hub\cite{dockerhub}, such as ubuntu:14.04, etc. The source code of dynamically link libraries in the images can be obtained after detecting the release versions of them. In addition, it is possible to replace the original glibc or other libraries in the image with the self-compiled ones. Therefore, building the mapping based on source code and binary analysis makes sense. A Program may leverage the embedded assemble code (e.g., the 'syscall' instruction) or the syscall() API or its own dynamically-linked libraries to invoke syscalls. For these programs, we can analyze the binary of the programs to extract the syscall list.

\begin{figure}
\centering
\includegraphics[width=0.35\textwidth]{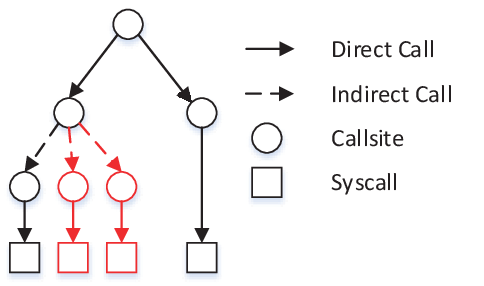}
\caption{Indirect calls enlarge the mapping.}\label{fig:indirect-call} 
\end{figure}

However, static analysis cannot precisely analyze some complex cases, such as indirect calls. As shown in Figure \ref{fig:indirect-call}, an indirect call could have multiple destinations, which are different callbacks of an operation. Due to the limitation of static analysis, there are some false positive destinations (red in the figure), which will never be executed in the normal execution paths. If static analysis is the only way to analyze the accessible syscall set, the false positive destinations will introduce extra syscalls, enlarging the attack surface. To overcome this challenge, a dynamic verification is needed to check if the indirect-call-related syscalls are really triggered by the secure path at runtime.

\subsection{System Overview}

\begin{figure*}
\centering
\includegraphics[width=0.85\textwidth]{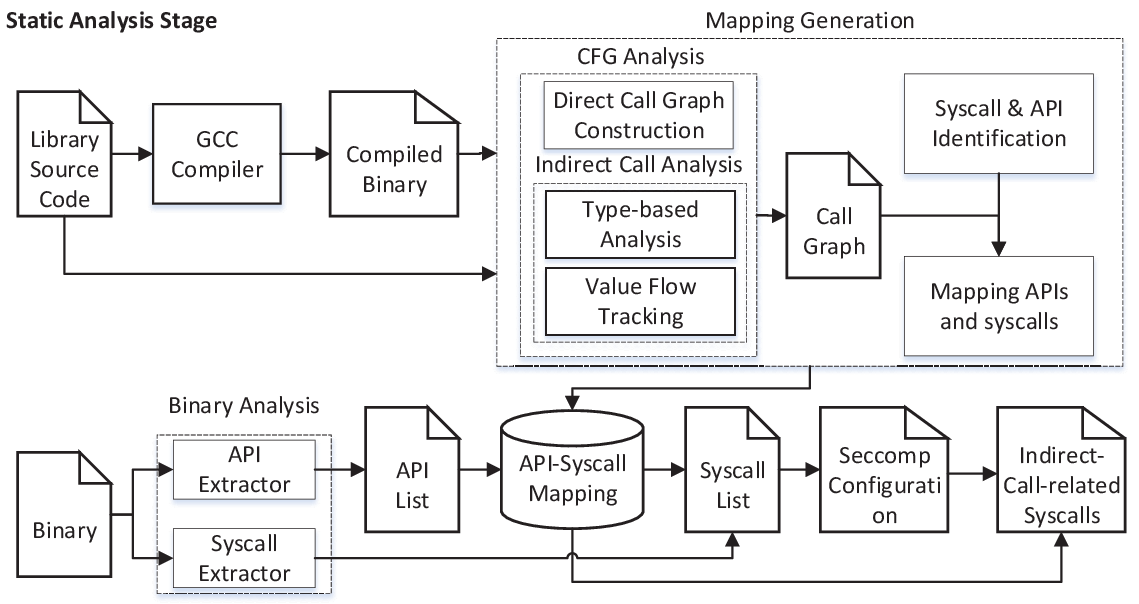}
\caption{The workflow of static analysis.} \label{fig:arch} 
\end{figure*}

There are two stages in our system: the static analysis stage and the dynamic verification stage. The static analysis stage analyzes the target binary and the related libraries to obtain the dependent syscall list and the indirect-call-related syscalls. The dynamic verification stage leverages the Seccomp to limit the syscall access of the target program, then it performs verification on suspicious syscalls to further reduce the attack surface.

The workflow of the static analysis stage is shown in Figure \ref{fig:arch}. The system can be mainly divided into two parts, including the mapping building part and the profile generation part. The mapping building part builds the mapping between the library APIs and syscalls. Then, the mapping is used to generate the Seccomp configuration. When a binary needs to be analyzed, the binary analysis module checks whether the binary invokes syscalls through the shared libraries or the embedded assembly code. If the shared libraries are used to invoke syscalls, the API extracting module can extract all of the dependent APIs. Based on the API-syscall mapping constructed by the mapping building part, the corresponding syscall list can be obtained. Finally, the Seccomp configuration can be generated based on the syscall list. If the binary invokes syscalls through the assembly code, the syscall extractor can analyze the assembly code to extract the code that invokes syscalls and then identify the corresponding syscall numbers.

The mapping between the library APIs and syscalls is constructed based on the combination of binary analysis and source code analysis. The source code of the library is compiled with debug information, which is used for direct function call graph (FCG) construction. By analyzing the disassemble instructions of the binary, the function start and end addresses can be identified. Next, the call instructions are used to construct the function call graph. The indirect calls are ignored in the binary analysis. 

Then, the indirect calls are analyzed by the source code analysis, since the source code contains more information about function calls. We leverage two approaches to analyze the indirect calls, including the address-taken function identification and the type-based function identification. These two approaches can build an over-approximate call graph. Finally, the whole FCG can be constructed by combing the direct FCG and indirect FCG together based on the debug information in the binary.

After constructing the call graph, the library APIs and the functions that invoke syscalls are identified. The external library APIs are the interfaces provided to user-space programs, which are the start nodes of call graphs. If an external API function can finally reach the functions that invoke syscalls, the API name and corresponding syscalls are added to the mapping. By analyzing all of the API functions, the comprehensive mapping can be obtained. 

During this process, we also collect the syscalls that are related with the indirect calls. From a library API, if the invocation path of a syscall contains an indirect call, this syscall is marked as an indirect-call-related syscall of the API. The indirect-call-related syscalls of a target program should be allowed, but some of them are false positives.

\begin{figure}
\centering
\includegraphics[width=0.45\textwidth]{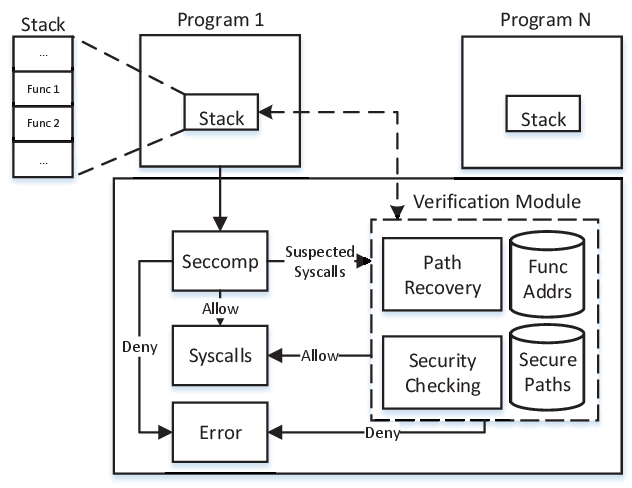}
\caption{The architecture of the dynamic verification module.}\label{fig:dynamic} 
\end{figure}

The dynamic verification module works in the kernel during the execution of the target program, whose architecture is shown in Figure \ref{fig:dynamic}. The syscall list extracted from the static analysis stage are used for the Seccomp configuration, so only these syscalls can be invoked by the target program. For the suspicious syscalls, the dynamic verification module hooks the related syscalls in the kernel to check if the invocation is caused by the secure invocation paths, which can detect false positives of the static analysis. Otherwise, the syscall is not included in the legal execution path and should be denied. 

Based on the observation that some syscalls that invoked by a library API only appear in rare cases, we propose another dynamic verification strategy, which collects the frequently invoked syscalls of a program and verifies other syscalls of the dependent syscall list dynamically. These syscalls make the accessible syscall set big. If all of the accessible syscalls are allowed by Seccomp, many vulnerable syscalls can be invoked by attackers through ROP or other attacks. But, if we can block and verify the invocation of these rarely invoked syscalls, the attackers can only invoke these syscalls through the fixed execution paths and logic within the service program, making the attacker more difficult to perform the attacks. The strategy minimizes the attack surface and leaves most of syscalls to be verified. Since most of the frequently invoked syscalls are allowed directly, the additionally introduced overhead will be low. Therefore, the suspicious syscalls can be the indirect-call-related syscalls or the rarely invoked syscalls.

\section{Systematic Static Analysis}\label{S:static-analysis}
The static analysis stage aims to construct the over-approximated mapping between library APIs and syscalls by analyzing the binaries and source code of the dependent libraries. The direct call graph is constructed by binary analysis. The indirect calls are analyzed based on the source code. Based on the full call graph, we can map the library APIs with the syscalls.

\subsection{Direct Call Graph Construction} 
We analyze the compiled binary of a dynamically-linked library to build the direct call graph. To that end, we disassemble the binary file of the target library and find all of the functions and their addresses. Then, all of the callq instructions are analyzed. If a callsite is a direct call, the target function's address is the operand of the callq instruction, and the corresponding function name will be in the comment. By analyzing the addresses, a direct call graph can be constructed. If the operand of a callq instruction is not a constant value, the callsite is an indirect call. Since the binary does not contain enough information of the indirect calls (e.g., the types of the parameters), we do not analyze the indirect calls using binary analysis. 

Compared with analyzing the source code, it is easier to analyze the binary for the call graph construction. That is because there are many library-specific implementation methods of different libraries' source code and the glibc is not supported for the Clang/LLVM compiler.
First of all, there are many macros in the c source files. When recognizing function statements, it is difficult to track the actual code corresponding to these macros. Besides, there are many types of macros, and some of them are deep nested macros, which are more difficult to analyze. Secondly, there are many low-level specific implementations in the c source files. For example, there are multiple definitions of the same function. It is not easy to determine which function is actually applied. Finally, it is often found that the called function name is an alias function name, which means we cannot find the corresponding function based on that name. Through our observation, there are many types of alias function names in the glibc. In different circumstances, an alias function name can be mapped to different functions, which is not easy to analyze.

Another problem of the source code analysis is that the analysis of the syscall invocation is library-specific. As described in the glibc wiki\cite{glibc-wiki}, there are three types of implementations for the syscall invocation. The first type is the assembly syscalls, which are translated from a list of names into an assembly wrapper that is then compiled using the syscall wrapper in the syscall-template.S. The list of syscalls that use wrappers is kept in the syscalls.list files. So the analyzer cannot find the syscall instructions of these syscalls in the source code. The second type is the macro syscalls. When system calls need to be called in the c source file, macros are used. There are many syscall-related macro definitions in the ``sysdeps/unix/sysdep.h" file, such as ``SYSCALL\_CANCEL". The macros are all called INTERNAL\_* and INLINE\_* and provide several variants to be used by the source code. The final type is the bespoke syscalls. In the glibc, some functions are implemented in the assembly code instead of the C language. For example, the implementation of the function syscall() is in the file ``sysdeps/unix/sysv/linux/x86\_64/syscall.S". To construct the function call graph, the assembly analysis is also needed. In contrast, these problems do not exist in binary analysis.

In summary, the problems of source code analysis are avoided by the binary analysis, which does not need to handle the complicated library-specific implementation details. But, it is not easy to analyze the indirect calls by using the binary analysis, that is because obtaining the number and types of the callsite parameters is difficult in binary analysis.

\subsection{Indirect Call Analysis}
An example of indirect call is illustrated in Listing 1. The upper disassembly code is a fragment of the disassembly code of the key\_call\_socket function, the source code of which is in /sunrpc/key\_call.c of glibc. From the disassembly code ``callq *(\%rax)" of the indirect callsite, we cannot obtain the destination. In the source code (Line 16), this indirect call is implemented through a macro "clnt\_call", which is defined in /sunrpc/rpc/clnt.c. This macro first takes the required function pointer from a structure, and then calls the function through the pointer. 

\begin{lstlisting}[caption=An example of indirect call.]
// The disassembly of snippet of key_call_socket function
11d5e2:	48 8b 47 08          	mov    0x8(%rdi),%rax
11d5e6:	6a 00                	pushq  $0x0
11d5e8:	6a 1e                	pushq  $0x1e
11d5ea:	4d 89 f1             	mov    %r14,%r9
11d5ed:	4d 89 e8             	mov    %r13,%r8
11d5f0:	4c 89 e1             	mov    %r12,%rcx
11d5f3:	48 89 ea             	mov    %rbp,%rdx
11d5f6:	48 89 de             	mov    %rbx,%rsi
11d5f9:	ff 10                	callq  *(%rax)

// Snippet of key_call_socket function in /sunrpc/key_call.c
if (clnt != NULL){
    wait_time.tv_sec = TOTAL_TIMEOUT;
    wait_time.tv_usec = 0;
    if (clnt_call (clnt, proc, xdr_arg, arg, xdr_rslt, rslt, wait_time) == RPC_SUCCESS)
        result = 1;
}

// The definition of clnt_call in /sunrpc/rpc/clnt.c
#define	clnt_call(rh, proc, xargs, argsp, xres, resp, secs)	\
	((*(rh)->cl_ops->cl_call)(rh, proc, xargs, argsp, xres, resp, secs))
\end{lstlisting}

For indirect calls, we analyze the source code to find the possible destinations of them, since the source code contains more information than the binary. We adopt the two-layer indirect call analysis of \cite{lu2019detecting} in our system, which first tracks the function pointers and then leverages type-based alias analysis\cite{niu2014modular,tice2014enforcing,farkhani2018effectiveness} to determine the destination functions. These approaches are implemented in the LLVM pass modules, but sysverify can only analyze the source code, because the glibc cannot be compiled by the Clang/LLVM compiler. The address-taken functions are identified by checking if a function's pointer is assigned to a variable or a structure field. Only the address-taken functions can be the destinations of indirect calls. Then, the type-based alias analysis is used to determine the possible destinations, which compares the number and types of the callsite's parameters and those of the possible callees. The functions with the same types of parameters are the possible destinations. Based on the indirect call analysis, the indirect function call graph can be constructed. The points-to analysis would help in addition to the address-taken function identification and the type-based function identification, but it is well known that the points-to analysis is limited in terms of accuracy and performance\cite{jeong2019razzer}. In addition, the time complexity of a typical points-to analysis (i.e., an Andersen analysis \cite{andersen1994program}) is O($n^3$), where n denotes the size of the program to be analyzed, it would take a very long time to analyze the entire glibc project. So, the points-to analysis is not applied in this work currently, and how to apply the points-to analysis is left in the future work.

\subsection{Generating the Mapping}
After constructing the direct and indirect function call graphs, sysverify combines them to construct a comprehensive function call graph. The dynamically-linked libraries are compiled with debug information, so the source code location of every function can be found in the binary code. Based on the source code location, the indirect function call graph can be integrated into the direct function call graph. In the comprehensive function call graph, sysverify can take a function as the starting point and perform a breadth-first search to know which functions can be involved in the control flow.

Next, the API function and syscalls are identified in the function call graph. The API function can be identified by the comments in the binary. For instance, the API functions of the glibc is marked as ``$<$API\_name@@GLIBC\_\_version$>$". The syscalls are invoked by the syscall instruction in the binary, which are easy to identify. The corresponding syscall number is stored in the RAX/EAX register, which can be analyzed by using the static data flow analysis. The details are described in Section \ref{s:sysname}.

The API-Syscall mapping is constructed by collecting the reachable syscalls of the API functions in the function call graph. However, the mapping is over-approximated, since the static analysis cannot identify the possible destinations of an indirect call precisely. To assist the dynamic verification, the syscalls introduced by the indirect calls are marked in this step by analyzing the indirect-call-related execution paths.

\section{Dynamic Verification}\label{s:dynamic}
After obtaining a full set of the dependent syscalls of a program, the dynamic verification module firstly leverages the Seccomp to limit the syscall invocations. For the allowed syscalls, it checks if the suspicious (i.e., indirect-call-related or rarely invoked) syscalls are invoked through the secure execution paths. Other allowed syscalls are executed without checking. The allowed syscalls refer to the dependent syscalls obtained by static analysis whose invocation paths contain only direct calls and which are invoked frequently. The secure execution paths refer to fixed execution paths and logic within the glibc from APIs to syscalls. In static analysis, if an invocation path of a syscall contains indirect calls, we cannot be sure that this syscall will be actually invoked by the API or is just a false positive caused by inaccurate indirect call analysis. If it is only a false positive, it enlarges the kernel attack surface. When the invocation path of a syscall matches the path obtained in our static analysis, the syscall is API-dependent and is considered to be secure; otherwise it may be invoked by attackers using attacks such as ROP. The execution path of a library API is recovered by analyzing the stack of the target program. If the extracted path matches one of the secure paths obtained by the static analysis stage, the invocation is considered as secure.

\subsection{Locating Functions in Memory}
To verify if the syscall invocation path is secure, we need to recover the invocation path from the extracted memory addresses in the stack. To that end, sysverify has to know the mapping between the functions obtained by the static analysis with their memory addresses. Since the  dynamic memory location cannot be obtained by the static analysis, we need to locate the library functions in the memory dynamically.

To find the memory location of the dynamically-linked libraries, we use a customized elf loader to load the corresponding libraries at runtime, which can output the memory locations of the loaded libraries. When a memory location is obtained, sysverify sends it to the kernel module.

Next, the function offsets within the library binary is obtained. It is not easy to recognize the code slice of a function in a compiled glibc binary\cite{meng2016binary}, due to many optimization technologies, such as code reuse, etc. So, we compile the library with the debug information. By using the debug information, we can know all of the function addresses in the binary. When the binary is loaded into the memory, the function addresses in memory can be calculated based on the base address of the library. With the set of the function addresses in memory, the dynamic verification module can map every instruction with the corresponding function.  

\subsection{Hooking Syscalls}
The dynamic verification module works as a kernel module intercepting the suspicious syscalls that are allowed by the Seccomp. The kernel module hooks the corresponding syscalls instead of the kernel entry point, so it only introduces overhead to the related syscalls. When an intercepted syscall is invoked, the module first determines if the invoking process is the target process by checking the CR3 value of it. If the invoking process is the analysis target, the verification module continues to check if the invocation path is secure. Otherwise, the syscall can execute directly after the comparison of the CR3 value. At this time, the overhead is only introduced by the value comparison, which is very low.  

\begin{figure}
\centering
\includegraphics[width=0.45\textwidth]{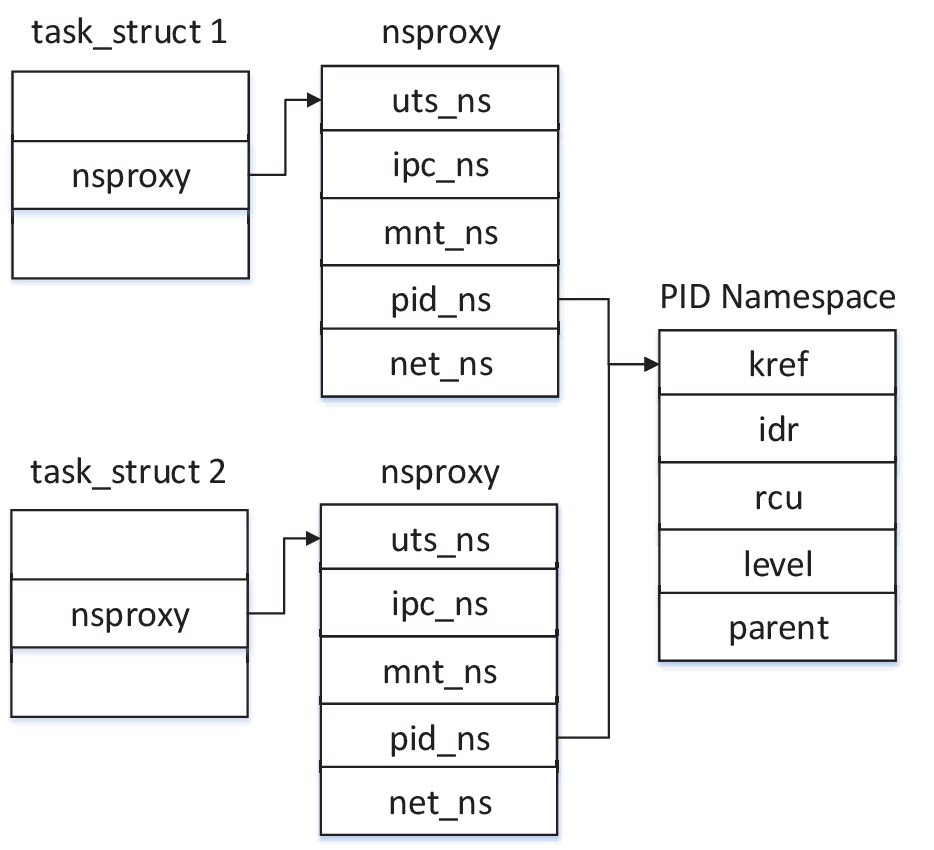}
\caption{Two processes share the same PID namespace.}\label{fig:ns} 
\end{figure}

For the container processes, the kernel module identifies the namespace fields of the corresponding task structure to determine if a process belongs to the target container. The Linux namespaces are used to isolate different containers in the same host. There are 6 different namespaces in Linux, including the PID namespace, the UTS namespace, the IPC namespace, the MNT namespace and the USER namespace. These namespaces are used to isolate different resources. Two processes with the same namespace can share the same resource. For instance, if two processes share the same PID namespace, they can see each other and the pids of them share the same space, as shown in Figure \ref{fig:ns}. When a container is initialized, the new namespaces are automatically created and assigned to the processes in the container. Processes in the same container have the same namespace, so the corresponding fields of the task structures are the same. When a target container is initialized, the corresponding namespace pointer is identified, which is used to identify the target processes at runtime.

\subsection{Recovering the Invocation Path}
The dynamic verification module reconstructs the invocation path by analyzing the stack of the target process. A library API usually executes many functions then invokes the syscall. When a function is called by the call instruction, the arguments and the return address are pushed into the stack. Therefore, the stack consists of addresses and data, and the top of it is stored in the RSP register. The library functions usually use the syscall instruction to invoke syscalls. This instruction switches the CPU to the kernel mode and stores the user-space RSP value in the task structure. Then, the kernel entry code executes different syscalls based on the syscall number. If a syscall needs to be verified, the injected kernel module reads the user-space stack and RIP of the target process from the kernel. The RSP value is first checked to see whether it locates in the stack range, because some ROP attackers usually use other memory regions as the malicious stack (e.g., the stack pivoting attacks) to enlarge the controllable buffer. The RIP is also checked to determine if the invocation is from the proper position. The stack is read from the top (i.e., the RSP value stored in the task structure). 

Next, the invocation path is reconstructed. The reconstruction algorithm is shown in Algorithm \ref{A:1}. For every value in the stack, if the value is in the range of the library addresses, it will be treated as an address. Then, the module identifies the function of it by matching the address with those of different functions. When the address reaches to the code segment of the target process, the module terminates. By analyzing all of the addresses in the stack, the execution path of the invocation can be reconstructed.

\begin{algorithm}[t]
\caption{Invocation Path Reconstruction}
\label{A:1}
\hspace*{0.02in} {\bf Input:}
Stack\_Content, Function\_Addresses(F\_A)\\
\hspace*{0.02in} {\bf Output:}
Execution\_Path
\begin{algorithmic}[1]
\For{Value \textbf{in} Stack\_Content}
    \If{Value \textbf{in} range(min(F\_A), max(F\_A))}
        \State func = Find\_Function (Value, F\_A)
        \State Execution\_Path.Insert(func)
    \ElsIf{Value \textbf{in} Code\_Segment}
            \State break
    \Else
        \State Pass
    \EndIf
\EndFor
\State \Return Execution\_Path
\end{algorithmic}
\end{algorithm}

The verification module matches the reconstructed execution path with the secure paths obtained from the static analysis. If the module cannot find the execution path from the secure paths, the invocation is not secure and will be denied. If the invocation is secure, it is approved, and the corresponding syscall will be added into the secure syscall set. Therefore, all of the suspicious syscalls are only analyzed once, the verification module does not introduce high overhead to the target process. 

\subsection{Discussion}
The dynamic verification module secures the suspicious syscalls by reconstructing and analyzing the invocation paths of them. Compared with the approaches only using static analysis to generate the syscall list, sysverify is more powerful.

However, if an attacker can tamper with the whole stack and overwrite the current user-space stack with the return addresses of a secure invocation path before the malicious syscall invocation, the verification module could be deceived. That is because the module reconstructs the invocation path based on the fake stack, so the invocation is considered the to be secure.

There are several challenges to perform this attack. The first one is the canary mechanism, which inserts some canary bits into the stack to detect stack overflow. If the attacker overwrites the bits, the process will be terminated. This mechanism limits the length of the controllable stack, so the attacker cannot fake the whole stack to cheat the verification module. There are some methods to bypass the canary mechanism, such as reading the canary bits and writing the bits during the stack overwriting. To overcome this kind of attack, we customize the dynamically-linked libraries and make it difficult to fake the stack. For the public releases of some popular libraries (e.g., glibc), the attackers can know the call relationship between functions and infer all of the function addresses based on a leaked memory address of the library. But, if the library is customized or recompiled with some specific flags, the attackers cannot know the function offsets and the call graph will be changed (e.g., disabling the inline functions). Since the attacker cannot obtain the binary of the customized glibc, it is difficult for them to fake the stack and remain undetected. In addition, there are many approaches to prevent the stack flow attacks\cite{duck2017stack,wang2018detect,lopez2019anatomy}. 

The second challenge is that faking the whole stack will affect the control flow of the attack. Attackers usually invoke some high privileged syscalls or exploit the vulnerabilities in some syscalls to perform  privilege escalation attacks\cite{CVE-2016-8655,CVE-2017-5123,CVE-2017-7308}. If the stack is overwritten with a secure execution path, the control flow will return to the corresponding path, which could make the process crash, and the attacker cannot continue the next step. 

Therefore, the dynamic verification module enhances the security of the kernel and makes attacks very difficult. 

Another limitation of sysverify is that it only supports docker containers currently. Other types of containers such as Kata, gVisor are based on virtual machines, and we leave how to apply sysverify to VM-based containers in the future work.

\subsection{Dynamic Tracking}
Besides verifying the indirect-call-related syscalls, we also propose another verification strategy, which only lets the frequently invoked syscalls execute directly, other syscalls are checked when they are invoked for the first time. In this strategy, more syscalls are checked without introducing high overhead, because only rarely invoked syscalls are intercepted. To collect the frequently invoked syscalls, we execute the target programs with different inputs for 100 times and leverage the strace to collect the invoked syscalls. These syscalls can be executed directly at runtime, other syscalls obtained by the static analysis will be checked by the verification module.

\section{Implementation}\label{s:implementation}
In this section, some important implementation details are described. The prototype system is implemented in the Ubuntu 20.04 (Linux 5.8) with the glibc v2.31.

\subsection{Compilation}
The glibc is compiled by the gcc compiler with the flags of ``-g -O2 -fdump-ipa-cgraph  -fdump-tree-cfg". By using these flags, the debug information will be added into the binary, and the cgraph and cfg files will be dumped during the compilation. These files can assist the source code analysis. The cgraph files record the alias function names. The cfg files are dumped after the control and data flow analysis, which represent the of control flow graph of each function.

\subsection{Mapping Syscall Names and Nums}\label{s:sysname}
After constructing the function call graph, the functions containing syscall invocation are analyzed to identify the invoked syscall name. The disassembly tool objdump is used to disassemble the binary with the option -d. In assembly code, the syscall invocation is based on the syscall instruction, and the syscall number is passed to the RAX/EAX register. To identify the corresponding syscall number, we explore the target binary to find out the methods to pass the syscall numbers to the target register. Through our observation, there are three approaches to pass the number, including: 1) passing a constant value to the RAX/EAX register; 2) passing the value through other registers; 3) passing the result of some numerical calculations to the register.

We leverage a backward data flow analysis to identify the syscall number starting from the syscall instruction. The instructions operating the RAX/EAX register are identified before the syscall instruction. Next, the instructions operating these registers are identified. The termination condition is the constant assignment, all of the extracted instructions are combined and analyzed together. The analysis is forward, which starts from the first instruction. When there is a constant passing to the RAX/EAX register, the constant is the syscall number. If the constant is passed to other register, this constant is hold, and the corresponding register is mark as the constant value until it reaches the RAX/EAX register. If there are some numerical operations, the calculation will be performed to further track the value in the register.   

After finding the syscall numbers, we use the syscall\_64.tbl file in the Linux source code to construct the relationship between the names of the system call numbers.

\subsection{Indirect-call analysis}
The indirect calls are analyzed by the source code analysis, and we leverage the compiling dumps to assist the analysis. Based on our observation, the callsites are easier to analyze with the cfg files, because these files are generated after the control flow and data flow analysis of the compiler. The cfg files represent the control flow graph of every function. In the cfg files, there are many code blocks in every function. Each code blocks are marked with the block labels, and they are connected by the goto statements. By analyzing all of the code blocks, the indirect calls can be identified by checking if the operand of a call statement is a variable. 

The address-taken functions are firstly identified. If a function is used as a variable, it is an address-taken function. To identify them, we analyze every statement in the cfg files to find the statements which operands are function names. For instance, passing a function to a variable as a pointer or passing a function to a parameter of a callsite. By extracting and collecting the function names, the address-taken function set is built.

The type-based alias analysis is used to find out the target destinations of an indirect call, which compares the number and types of the parameters with those of the possible functions. To this end, we collect the parameters of every possible function, and then count the number and types of them. It is easy to collect the types of a function's parameters. In contrast, collecting the types of a callsite's parameters is more challenging, because the parameters of a callsite are variables without the type information. To analyze the types of the them, the backward data flow analysis is used based on the observation that the types of the variables are located in the same function. Some variables are declared in the same function, so the types of them can be collected from the declarations. Other variables are passed from the parameters of the current function, which types can be collected from the definition of the function. With the type information, the possible callees can be identified.

\subsection{Hooking Syscalls}
The kernel module hooks some syscalls and performs the security verification in the kernel. When the system switches to the kernel mode, the kernel entry point is executed. It locates the corresponding syscall based on the sys\_call\_table and the syscall number. To hook the syscalls, sysverify replaces the addresses of the syscalls to be monitored in the sys\_call\_table with the security module. As a result, when the syscall is executed, the security module is executed first. After the verification, the original syscall will be executed if it is secure.

The security module uses the current task structure and the task\_pt\_regs function to obtain the register information of the user-space task. The CR3 can be obtained by the function read\_cr3\_pa, which is used to check if the current process is the target process to be monitored. The user-space stack is read by the function copy\_from\_user based on the value of RSP. Based on the content, the verification can be performed.

\section{Evaluation}\label{s:evaluation}
We evaluated sysverify with a set of 100 popular applications in Linux, including utility tools and applications extracted from the most popular Docker containers images. We evaluate the effectiveness and performance of our system.

This paper mainly focuses on C/C++ programs, so we collect the most popular utility tools and applications written in C/C++ for the evaluation. Utility tools are used in high frequency, so we collect the 80 most common of them in the directories in Linux, such as ps, ls, etc. We also collect programs in the Docker container images. We select and analyze 20 most downloaded C/C++ programs running in containers. To find out which programs are executed in containers, we run these Docker images respectively and track the executed binaries. Finally, the list of target binaries is extracted.

\subsection{Effectiveness}
We evaluate if sysverify can filter unused syscalls without affecting the execution of target programs. Then, it is compared with the related work. Finally, we analyze if sysverify can prevent CVE-related syscalls to mitigate the CVEs.

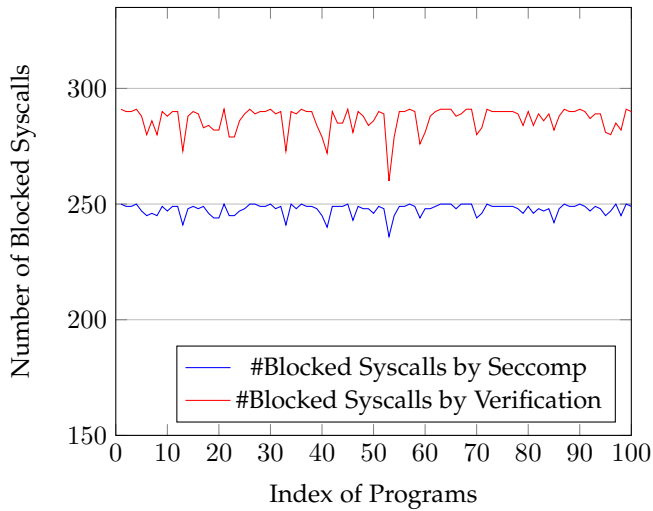
\begin{figure}
    \centering
    \begin{tikzpicture}
    \begin{axis}[
			xlabel = Index of Programs,
			xmin = 0,
			xmax = 100,
			ymin = 150,
			ymax = 335,
			ylabel= Number of Blocked Syscalls,
			ymajorgrids = true,
			legend pos=south east,
			xtick= {0, 10, 20, 30, 40, 50, 60, 70, 80, 90, 100},
			]
    \addplot[blue] plot coordinates {
(1, 250) (2, 249) (3, 249) (4, 250) (5, 247) (6, 245) (7, 246) (8, 245) (9, 249) (10, 247) 
		(11, 249) (12, 249) (13, 241) (14, 248) (15, 249) (16, 248) (17, 249) (18, 246) (19, 244) (20, 244) 
		(21, 250) (22, 245) (23, 245) (24, 247) (25, 248) (26, 250) (27, 250) (28, 249) (29, 249) (30, 250) 
		(31, 248) (32, 249) (33, 241) (34, 250) (35, 248) (36, 250) (37, 249) (38, 249) (39, 248) (40, 245) 
		(41, 240) (42, 249) (43, 249) (44, 249) (45, 250) (46, 243) (47, 249) (48, 248) (49, 248) (50, 246) 
		(51, 249) (52, 248) (53, 236) (54, 245) (55, 249) (56, 249) (57, 250) (58, 249) (59, 244) (60, 248) 
		(61, 248) (62, 249) (63, 250) (64, 250) (65, 250) (66, 248) (67, 250) (68, 250) (69, 250) (70, 244) 
		(71, 246) (72, 250) (73, 249) (74, 249) (75, 249) (76, 249) (77, 249) (78, 248) (79, 246) (80, 249) 
		(81, 246) (82, 248) (83, 247) (84, 248) (85, 242) (86, 248) (87, 250) (88, 249) (89, 249) (90, 250) 
		(91, 249) (92, 247) (93, 249) (94, 248) (95, 245) (96, 247) (97, 250) (98, 245) (99, 250) (100, 249) 
    };
    \addlegendentry{\#Blocked Syscalls by Seccomp}
    \addplot[red] plot coordinates {
(1, 291) (2, 290) (3, 290) (4, 291) (5, 288) (6, 280) (7, 286) (8, 280) (9, 290) (10, 288) 
		(11, 290) (12, 290) (13, 273) (14, 288) (15, 290) (16, 289) (17, 283) (18, 284) (19, 282) (20, 282) 
		(21, 291) (22, 279) (23, 279) (24, 286) (25, 289) (26, 291) (27, 289) (28, 290) (29, 290) (30, 291) 
		(31, 289) (32, 290) (33, 273) (34, 290) (35, 289) (36, 291) (37, 290) (38, 290) (39, 284) (40, 279) 
		(41, 272) (42, 290) (43, 285) (44, 285) (45, 291) (46, 281) (47, 290) (48, 288) (49, 284) (50, 286) 
		(51, 290) (52, 289) (53, 260) (54, 279) (55, 290) (56, 290) (57, 291) (58, 290) (59, 276) (60, 281) 
		(61, 288) (62, 290) (63, 291) (64, 291) (65, 291) (66, 288) (67, 289) (68, 291) (69, 291) (70, 280) 
		(71, 283) (72, 291) (73, 290) (74, 290) (75, 290) (76, 290) (77, 290) (78, 289) (79, 284) (80, 290) 
		(81, 284) (82, 289) (83, 286) (84, 289) (85, 282) (86, 288) (87, 291) (88, 290) (89, 290) (90, 291) 
		(91, 290) (92, 287) (93, 289) (94, 289) (95, 281) (96, 280) (97, 285) (98, 282) (99, 291) (100, 290)
    };
    \addlegendentry{\#Blocked Syscalls by Verification}
    \end{axis}
    \end{tikzpicture}
    \caption{The statistics of the syscall dependency analysis on every program.}\label{fig:static-result-everyp-program}
\end{figure}

We first leverage the static analysis to obtain the over-approximated dependent syscall lists of the target programs. The corresponding Seccomp configurations and indirect-call-related syscalls are generated automatically. The statistics of the experiment is illustrated in Table \ref{T:statistics }. Every syscall is an entry point to the kernel functions, according to the measurement of Nabla\cite{nabla}, reducing the number of accessible syscalls can shrink the accessible kernel functions. The analysis results are shown in Figure \ref{fig:static-result-everyp-program}. The figure shows the number of system calls blocked for each program through static analysis. The red line is the blocked syscalls of sysverify with dynamic verification, and the blue line is the syscalls blocked by Seccomp. The 
blue line shows that there are at least 236 syscalls (out of 335) disabled from the target programs in the worst case. There are 247.71 syscalls can be disabled on average. In contrast, the Docker only disables 49 syscalls by default. According to the analysis of the indirect calls, there are 39.03 indirect-call-related syscalls in the dependent syscall list of a program on average. These syscalls can be false positives, verifying these syscalls can further reduce the attack surface.

\begin{table}
  \centering
  \caption{The statistics of the experiment.}\label{T:statistics }
  \begin{tabular}{ll}
     \hline
     Category & Number  \\
     \hline
	Direct calls&	12,739\\
    Indirect calls&	1,865\\
    Allowed syscalls on average&	87.29\\
    Disable syscalls on average	&247.71\\
	 \hline
   \end{tabular}
\end{table}

Next, we execute the target programs with the corresponding Seccomp configurations one by one using the zero-code-Seccomp approach, which does not need to modify the source code of the target programs and only injects policies into the target programs. During the execution, the dynamic verification module works in two different verification strategies respectively. The first one is to only intercept and analyze the indirect-call-related syscalls. The other one is to intercept the syscalls that are invoked in low frequency. For the second strategy, we run and track every program before the test to collect the invoked syscalls dynamically. This collection involves the syscalls that invoked in high frequency. To enlarge the set, we test the target programs using different inputs. The number of collected syscalls is 14.07 per program on average. The dynamically invoked syscalls only occupy 16.09\% of the over-approximated syscalls on average. The result means that there are lots of syscalls obtained by the static analysis will not be executed in the real execution. Compared with the related work \cite{ghavamnia2020confine} and \cite{demarinis2020sysfilter}, the verification of indirect-call-related or rarely invoked syscalls further reduces the kernel attack surface.

During the execution, we first check if the target programs can execute properly with sysverify, then find out how many indirect-call-related or rarely invoked syscalls are invoked. After testing every program, we find that all of these programs can execute properly, which proves the robustness of sysverify. Among the 100 programs, none of them invoke the indirect-call-related syscalls. Based on this result, we can find that most of the indirect-call-related syscalls will not be invoked. That is because some of them are false positives, and some of them are only invoked in occasional conditions. Therefore, verifying these syscalls can further shrink the attack surface. When sysverify is configured to intercept rarely invoked syscalls, there are some syscalls triggering the verification. For some complex programs, this number is up to 7 (MySQL). But for some small programs (e.g., ls) there are no newly invoked syscalls. The verification will introduces overhead to the target program, the performance measurement is performed in Section \ref{S:performance}. 

\subsection{CVE Mitigation}
Some syscalls can be exploited to perform privilege escalation or other attacks\cite{CVE-2016-8655,CVE-2017-5123,CVE-2017-7308}, so we measure how many CVEs can be mitigated by removing the unused syscalls. To that end, we first crawl the CVE website \cite{cve-website} and collect the kernel functions of the related CVEs. Then, we map the CVEs with syscalls by building the call graphs of the kernel functions with KIRIN\cite{zhang2019pex}, which can map the kernel functions with the syscalls. Through the mapping we can discover how many CVEs can be mitigated by the syscall limitation. 

The top 20 syscalls and corresponding CVEs filtered by the sysverify are shown in Table \textcolor{blue}{\ref{T:cve}}. There are totally 87 CVEs mitigated by the Seccomp configuration of the static analysis. Among them, 50 CVEs are removed from all of the analyzed programs. Besides the static analysis, the dynamic verification can further reduce the CVEs. By verifying the suspicious syscalls, up to 8 new CVEs can be mitigated. 

From the dynamic tracking results, the rarely invoked syscalls are invoked in very low frequency. So, when sysverify intercepts the rarely invoked syscalls, most CVE-related syscalls are not executed. Through verifying the rarely invoked syscalls, 71 CVEs can be mitigated on average. Compared with Confine \cite{ghavamnia2020confine}, the dynamic verification can effectively isolate more (i.e., 9.2\% by verifying the indirect-call-related syscalls and 81\% by verifying the rarely invoked syscalls) CVEs from the programs. The CVEs mitigated by the dynamic verification is related with how many suspicious syscalls invoked by the tested programs, so this result may change according to different programs.

\begin{table*}
  \centering
  \caption{Top 20 syscalls and the related CVEs mitigated by sysverify.}\label{T:cve}
  \begin{tabular}{ccccc}
     \hline
     Syscalls & Number of CVEs & Representative CVEs \\
     \hline
	ioctl & 29 & CVE-2016-0723,CVE-2009-1192,CVE-2015-0275 \\
	execveat & 10 & CVE-2015-3339,CVE-2010-4243,CVE-2012-4530 \\
	keyctl & 8 & CVE-2016-0728,CVE-2015-7550,CVE-2009-3624 \\
	ptrace & 5 & CVE-2009-1527,CVE-2019-13272,CVE-2014-9870 \\
	add\_key & 4 & CVE-2016-0728,CVE-2017-15274,CVE-2015-8539 \\
	mount & 3 & CVE-2014-5207,CVE-2014-5206,CVE-2008-2931 \\
	unshare & 2 & CVE-2013-1858,CVE-2013-4205 \\
	waitid & 2 & CVE-2017-14954,CVE-2017-5123 \\
	request\_key & 2 & CVE-2016-0728,CVE-2017-7472 \\
	rt\_tgsigqueueinfo & 2 & CVE-2013-2141,CVE-2011-1182 \\
	epoll\_ctl & 2 & CVE-2012-3375,CVE-2011-1082 \\
	move\_pages & 2 & CVE-2010-0415,CVE-2017-14140 \\
	shmctl & 2 & CVE-2009-0859,CVE-2010-4072 \\
	perf\_event\_open & 2 & CVE-2015-9004,CVE-2017-6001 \\
	clock\_nanosleep & 1 & CVE-2018-13053 \\
	semget & 1 & CVE-2015-7613 \\
	semctl & 1 & CVE-2010-4083 \\
	name\_to\_handle\_at & 1 & CVE-2012-6549 \\
	epoll\_pwait & 1 & CVE-2011-1082 \\
	fremovexattr & 1 & CVE-2011-1090 \\
	 \hline
   \end{tabular}
\end{table*}

\subsection{Performance}\label{S:performance}

The dynamic verification can introduce overhead to the target programs. To measure the overhead, we first run the target programs without verification for 100 times and record the execution time. Then, we test them using the same inputs under the dynamic verification with two different strategies. From the results, the dynamic verification introduces 1\% overhead on average in both strategies. The overhead is caused by three reasons, including comparing the CR3 value, reading the user-space stack and the function flow construction. For every intercepted syscall, the CR3 of the invoking process is compared. So, this is the major overhead of sysverify. We measure this overhead specifically by comparing the execution time of a syscall with and without CR3 interception respectively. We select the OPEN syscall as the analysis target. We invoke this syscall for 1,000 times and the execution time is 0.27ms. After adding the comparison of the CR3 value, the time is 0.28ms. Therefore, the overhead is less than 1\%.

Besides the major overhead, the overhead of sysverify is based on the frequency of the invocation path analysis. That is because the overhead of reading the user-space stack and the function flow construction is introduced only when the syscall invocation should be analyzed. But the newly invoked syscall is only checked once, so the overhead of sysverify is very low. 

We leverage the LMBench to test the overall overhead of sysverify when monitoring the rarely invoked syscalls, which can introduce the highest overhead. We select three benchmarks (e.g., syscall, block I/O and network I/O) to measure the overhead. The performance without sysverify is firstly measured, then we measure the different benchmark with sysverify intercepting the rarely invoked syscalls, which can introduce the highest overhead. From the result, we found that the overhead is not measurable, which means the overhead is smaller than the performance fluctuations. 

\section{Related Work}\label{s:related-work}
\subsection{VM-based Containers}
There are some VM-based containers proposed to further enhance the container isolation. The unikernel \cite{madhavapeddy2014unikernels} is based on the concept of the micro-kernel, using the library OS to compile the kernel part and the program that the application depends on into a virtual machine image. The image does not distinguish between the user layer and the kernel layer, and the operating system is directly integrated into the user program. The host can start the image in a virtual machine. This lightweight virtual machine has the characteristics of fast startup and small memory footprint. However, all applications need to be recompiled, and the lack of distinction between user space and kernel space reduces the security of it.

The Kata Container \cite{kata} is a hardware-assisted-virtualization-based container. In 2017, Intel Clear Containers\cite{clearcontainers} and Kata Containers were merged to use lightweight virtual machines to improve the isolation of containers. In Kata Containers, every container runs in a separate virtual machine. The Kata Container utilizes the Intel Clear Containers technology to make the virtual machine kernel lighter, enabling fast virtual machine startup and minimizing the resource usage. However, the Kata Container faces several problems. First, virtual machines occupy more resources than the Docker containers. Second, the initializing time of a Kata Container is longer than that of a Docker container.

The gVisor\cite{gvisor} is a para-virtualization-based container, which is developed by Google based on the go language, using para-virtualization to isolate containers. The runtime of gVisor is called runsc, which consists of two parts: Sentry and Gofer. Sentry is a virtual lightweight operating system kernel, which is responsible for handling all syscalls for containers. Sentry can simulate most syscalls, reducing the host's attack surface. However, some syscalls still cannot be simulated and need to be completed by the host operating system. Gofer is a proxy program of the container file system, which forwards all I/O requests of the container to the host. Since the gVisor also relies on the host's operating system, its isolation is weaker than that of full virtualization (e.g., Kata Containers). Moreover, the gVisor cannot implement some syscalls, so it does not support as many programs as the Docker containers.

In summary, VM-based containers face the problems of performance and versatility.
\subsection{Container Security}
Container security \cite{sultan2019container,tomar2020docker,bui2015analysis,yu2019survey,watada2019emerging} is a hot topic in cloud computing. \cite{combe2016docker} compares containers with virtual machines. Container is more light-weight, but it also raises security risks due to the weak isolation between containers and the host OS kernel. In addition, the ecosystem of containers can also introduce new security risks, such as vulnerable images.

\textbf{Container isolation.} Containers share the same kernel with the host, so the isolation is weaker than that of virtual machine. \cite{gao2019houdini} exploits the mechanism of cgroups and performs the cgroup escape attack, which can break the resource limitation of cgroups. Bastion \cite{nam2020bastion} describes the weaknesses of container network isolation and performs cross-container attack. \cite{gao2017containerleaks,gao2018study} present the isolation of the proc file system is weak for container scenario, and one container can perform some side-channel attacks to obtain the information of other containers in the same host. SCONE\cite{arnautov2016scone} leverages the Intel SGX to protect docker containers from external malicious attacks and untrusted cloud hosts. 

To enhance the isolation,  \cite{lei2017speaker,wan2017mining,barlev2016secure,ghavamnia2020confine} limit the number of syscalls that can be accessed by containers. Besides, \cite{ghavamnia2020temporal} identifies different service stages of some applications. Based on the observation that an application invokes different syscalls in different service stages (e.g., the initialization stage and the servicing stage), it applies different Seccomp configurations to the application according to its execution status. \cite{jian2017defense} checks the status of Linux namespaces to detect the container escape attacks. But, these approaches face the problems of the false positives introduced by static analysis and the unsystematic analysis of different libraries.

\textbf{Image security.} \cite{gummaraju2015over,henriksson2017static,shu2017study} have shown that 30\%-90\% of container images have security issues. \cite{lin2018measurement} and \cite{kabbe2017security} tested the possible vulnerabilities in the containers and explored the corresponding attack methods. \cite{mohallel2016experimenting} uses common applications such as MySQL and Apache to compare the security of containers and traditional servers. The results show that the overall security of the containers is worse due to some security vulnerabilities in the container images. \cite{abbott2017security} proposed a method for measuring the security of container images.

Researchers have proposed some methods to secure container images. \cite{martin2018docker} analyzes some security vulnerabilities in containers, and explores attack and protection methods. \cite{bila2017leveraging} can respond to anomalies after discovering security issues in container images, which is based on the Kubernetes to perform periodic vulnerability scanning on containers. After discovering anomalies, it can terminate container execution and recompile the container image to eliminate security risks.

\textbf{Monitoring containers.} \cite{watts2019insight} can obtain and analyze the information of the docker engine, docker instances and the host OS. \cite{bacis2015dockerpolicymodules} lets the image maintainers provide the SELinux security policies to enhance the security of the container by extending the Dockfile format. Therefore, different dockers can be configured by different SELinux policies, which can improve the security of the Docker container. \cite{gantikow2020container} analyzes the invoked syscalls to detect anomaly behaviors in containers. \cite{tien2019kubanomaly} detects abnormal behaviors by using the deep learning technology. 

\subsection{Program Debloating}
Program debloating is a way to reduce the attack surface by shrinking the code size of a target program or shared library. Piece-wise debloating \cite{quach2018debloating} leverages a customized binary loader to load only the required parts of the dynamic libraries by embedding the dependency information into the programs. The unused parts of the libraries are replaced. Nibbler\cite{agadakos2019nibbler} extracts the function call graphs of the target binaries and dependent libraries, then it removes the unused code. Razor\cite{qian2019razor} leverages dynamic tracking to extract the function call graphs. LibFilter\cite{shteinfeldlibfilter} can remove the unused functions in the dynamically-linked libraries. Shredder\cite{mishra2018shredder} can further limits the arguments of the dependent library APIs.

\section{Conclusion}\label{s:conclusion}
This paper proposes sysverify, which can shrink the kernel attack surface systematically by using the combination of static analysis and dynamic verification. The static analysis stage builds the over-approximated API-syscall mapping by analyzing the dependent shared libraries. The analysis leverages binary analysis to construct the direct function call graph and source code analysis to construct the indirect function call graph, which is more systematic than the approaches only analyzing the binaries or source code. The suspicious (i.e., indirect-call-related or rarely invoked) syscalls are verified by the dynamic verification module, which can reconstruct and analyze the invocation paths to further secure the kernel from the false positive syscalls introduced by the static analysis. The experimental results show that sysverify can remove the unused syscalls from the target programs and mitigate the CVEs included in the corresponding syscalls with low overhead. It is easy to extend sysverify to support the programs written in other languages by constructing the corresponding API-syscall mapping, and we leave it in future work.

\ifCLASSOPTIONcompsoc
  \section*{Acknowledgments}
\else
  \section*{Acknowledgment}
\fi

This work was supported by the National Key R\&D Program of China (No.2021YFB2012402) and the National Natural Science Foundation of China under Grants NO. 61872111.

\ifCLASSOPTIONcaptionsoff
  \newpage
\fi

\bibliographystyle{IEEEtran}
\bibliography{sec}

\end{document}